\documentclass[twocolumn,prl,superscriptaddress,amsmath]{revtex4-1}

\usepackage{graphicx}
\usepackage{hyperref} 

\begin{document}

\title
{Feedback-Induced Quantum Phase Transitions Using Weak Measurements}

\author{D. A. Ivanov}
\affiliation{Department of Physics, St. Petersburg State University, St. Petersburg, Russia}
\author{T. Yu. Ivanova}
\affiliation{Department of Physics, St. Petersburg State University, St. Petersburg, Russia}
\author{S. F. Caballero-Benitez}
\affiliation{Instituto de F\'isica, Universidad Nacional Aut\'onoma de M\'exico, Ciudad de M\'exico, M\'exico}
\author{I. B. Mekhov}
\email{Igor.B.Mekhov@gmail.com}
\affiliation{Department of Physics, St. Petersburg State University, St. Petersburg, Russia}
\affiliation{Department of Physics, University of Oxford, Oxford, United Kingdom}
\affiliation{SPEC, CEA, CNRS, Universit\'{e} Paris-Saclay, CEA Saclay, Gif-sur-Yvette, France}

\begin{abstract}
We show that applying feedback and weak measurements to a quantum system induces phase transitions beyond the dissipative ones. Feedback enables controlling essentially quantum properties of the transition, i.e., its critical exponent, as it is driven by the fundamental quantum fluctuations due to measurement. Feedback provides the non-Markovianity and nonlinearity to the hybrid quantum-classical system, and enables simulating effects similar to spin-bath problems and Floquet time crystals with tunable long-range (long-memory) interactions.
\end{abstract}

\maketitle

The notion of quantum phase transitions (QPT) \cite{SachdevBook} plays a key role not only in physics of various systems (e.g. atomic and solid), but affects complementary disciplines as well, e.g., quantum information and technologies \cite{Osterloh}, machine learning \cite{Nieuwenburg} and complex networks \cite{Halu2013}. In contrast to thermal transitions, QPT is driven by quantum fluctuations existing even at zero temperature in closed systems. Studies of open systems advanced the latter case: the dissipation provides fluctuations via the system-bath coupling, and the dissipative phase transition (DPT) results in a nontrivial steady state \cite{Kessler2012,Daley}. 

Here we consider an open quantum system, which is nevertheless not a dissipative one, but is coupled to a classical measurement device. The notion of fundamental quantum measurement is broader than dissipation: the latter is its special case, where the measurement results are ignored in quantum evolution \cite{Wiseman}. We show that adding the measurement-based feedback can induce phase transitions. Moreover, this enables controlling quantum properties of the transition by tuning its critical exponent. Such a feedback-induced phase transition (FPT) is driven by fundamentally quantum fluctuations of the measurement process, originating from the incapability of any classical device to capture the superpositions and entanglement of quantum world.

Feedback is a general idea of modifying system parameters depending on the measurement outcomes. It spreads from engineering to contemporary music, including modeling  the Maxwell demon \cite{Murch2018,Masayuma2018,Koski2015} and reinforcement learning \cite{Petruccione}. Feedback control has been successfully extended to quantum domain \cite{Wiseman,HammererRMP, HaukePRA2013,HuardFB2016, Hacohen2016, HarocheBook,Mabuchi2004, Sherson2015,Sherson2016, Hush2013,Bouchoule2017,Hammerer2016PRA, Thomsen2002,St-K2012,Vuletic2007,Ivanova2014,Wallentowitz2004,Ivanova2016} resulting in quantum metrology aiming to stabilize nontrivial quantum states and squeeze (cool) their noise. The measurement backaction typically defines the limit of control, thus, playing an important but negative role \cite{Ivanova2005}. In our work, we shift the focus of feedback from quantum state control to phase transition control, where the measurement fluctuations drive transition thus playing an essentially positive role in the process as a whole.  

\begin{figure}
\includegraphics[clip, trim=0cm 12.5cm 11.8cm 0cm, width=0.45\textwidth]{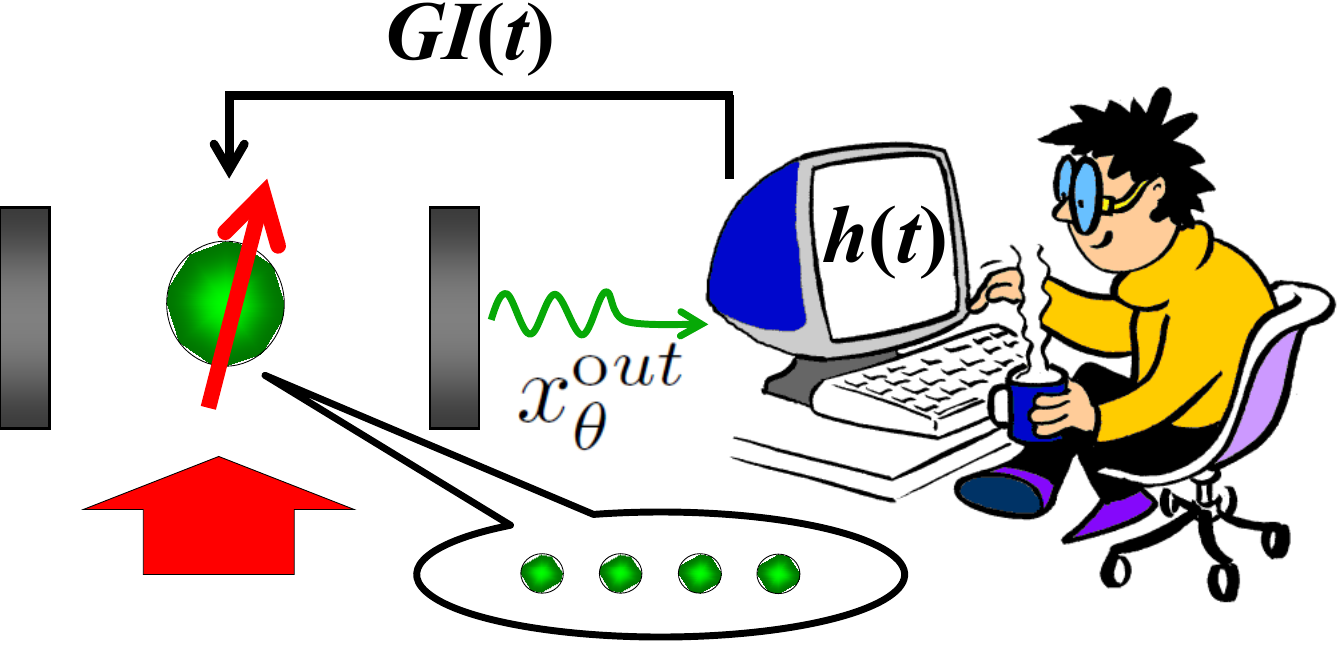}
\caption{\label{setup} Setup (details for a BEC system are given in \cite{Suppl}). Quantum dipoles (possibly, a many-body system) are illuminated by probe. Scattered light is measured and feedback acts on the system, providing non-Markovianity, nonlinearity, and noise, necessary for phase transition. Importantly, the feedback response $h(t)$ can be digitally tuned.} 
\end{figure}

Hybrid systems is an active field of quantum technologies, where various systems have been already coupled \cite{Kurizki2015}: atomic, photonic, superconducting, mechanical, etc. The goal is to use advantages of various components. In this sense, we address a hybrid quantum-classical system, where the quantum system can be a simple one providing the quantum coherence, while all other properties necessary for tunable phase transition are provided by the classical feedback loop: nonlinear interaction, non-Markovianity, and fluctuations. 

We show that FPT leads to effects similar to particle-bath problems (e.g. spin-boson, Kondo, Caldeira-Leggett, quantum Browninan motion, dissipative Dicke models) describing very different physical systems from quantum magnets to cold atoms \cite{breuer2002,Leggett1987,LeHur2008, DomokosPRL2015, DomokosPRA2016,Scarlatella2016,Plenio2011}. While tuning quantum baths in a given system is a challenge, tuning the classical feedback is straightforward, which opens the way for simulating various systems in a single setup. This raises questions about quantum-classical mapping between Floquet time crystals \cite{Sacha2017,Eckardt2017} and long-range interacting spin chains. Our model is directly applicable to many-body systems, and as an example we consider ultracold atoms in a cavity. Such a setup of many-body cavity QED (cf. for review \cite{Mekhov2012,ritsch2013}) was recently marked by experimental demonstrations of superradiant Dicke \cite{EsslingerNat2010}, lattice supersolid \cite{EsslingerNature2016, Hemmerich2015}, and other phase transitions \cite{LevPRL2018, Zimmermann2018}, as well as theory proposals \cite{Caballero2015,Caballero2015a, Rogers2014,Morigi2010,Niedenzu2013, Gopalakrishnan2009,Kollath2016,piazza2015self,Diehl2013,DomokosPRL2015, DomokosPRA2016}. Nevertheless, effects we predict here require to go beyond the cavity-induced autonomous feedback \cite{St-K2017}.

{\it Model.-} Consider $N$ two-level systems (spins, atoms, qubits) coupled to a bosonic (light) mode, which may be cavity-enhanced (Fig. 1). The Hamiltonian then reads
\begin{eqnarray}\label{Hamiltonian}
H=\delta a^\dag a +\omega_R S_z +\frac{2}{\sqrt{N}}S_x [g(a+a^\dag)+GI(t)],
\end{eqnarray}
which without the feedback term $GI(t)$ is the standard cavity QED Hamiltonian \cite{ScullyBook} describing the Dicke (or Rabi) model \cite{DomokosPRL2015, DomokosPRA2016} in the ultra-strong coupling regime \cite{HuardUltraS2018, Ustinov2017, Yoshihara2017, Forn2017} (without the rotating-wave approximation). Here $a$ is the annihilation operator of light mode of frequency $\delta$, $S_{x,y,z}$ are the collective operators of spins of frequency $\omega_R$, $g$ is the light-matter coupling constant. The Dicke model was first realized in Ref. \cite{EsslingerNat2010} using a Bose-Einstein condensate (BEC) in a cavity, and we relate our model to such experiments in Ref. \cite{Suppl}. Our approach can be readily applied to many-body settings as $S_x$ can represent various many-body variables \cite{Elliott2015,Kozlowski2015PRA,Kozlowski2017}, not limited to the sum of all spins: e.g., fermion or spin (staggered) magnetization \cite{Mazzucchi2016PRA, Mazzucchi2016SciRep,LandiniPRL2018,LevPRL2018} or combinations of strongly interacting atoms in arrays, as in lattice experiments \cite{EsslingerNature2016, Hemmerich2015}.  

The feedback term $GI(t)$ has a form of the time-dependent operator-valued Rabi frequency rotating the spins ($G$ is the feedback coefficient and $I(t)$ is the control signal). We consider detecting the light quadrature $x_\theta^{\text out}(t)$ ($\theta$ is the local oscillator phase) and define $I(t)=\sqrt{2\kappa}\int_0^t h(t-z)\mathcal{F}[x_\theta^{\text out}(z)]dz$. Thus, the classical device continuously measures $x_\theta^{\text out}$, calculates the function $\mathcal{F}$, integrates it over time, and feeds the result back according to the term $GI(t)$. In BEC~\cite{Suppl}, the quasi-spin levels correspond to two motional states of atoms, and coupling of feedback to $S_x$ is achieved by modifying the trapping potential~\cite{Suppl}. Various forms of the feedback response $h(t)$ will play the central role in our work. The input-output relation \cite{walls2008quantum} gives $x_\theta^{\text out}=\sqrt{2\kappa}x_\theta-f_\theta/\sqrt{2\kappa}$, where the intracavity quadrature is $x_\theta=(a e^{-i\theta}+a^\dag e^{i\theta})/2$ and $\kappa$ is the cavity decay rate. The quadrature noise $f_\theta=(f_a e^{-i\theta}+f_a^\dag e^{i\theta})/2$ is defined via the Markovian noise operator $f_a$ [$\langle f_a(t+\tau)f_a(t)\rangle=2\kappa \delta(\tau)$] in the Heisenberg-Langevin equation:
\begin{eqnarray}\label{Heis}
\dot{a}=-i\delta a -i\frac{2g}{\sqrt{N}}S_x-\kappa a +f_a.
\end{eqnarray}

{\it Effective feedback-induced interaction.-} An illustration that feedback induces effective nonlinear interaction is used in quantum metrology \cite{Thomsen2002} for a simple cases such as $I(t)\sim x_\theta^{\text out}$. One sees this, if light can be adiabatically eliminated from Eq.~(\ref{Heis}), $a\sim S_x$. Then the effective Hamiltonian, giving correct Heisenberg equations for spins, contains the term $S_x^2$ leading to spin squeezing \cite{Thomsen2002} [cf. Eq. (\ref{Hamiltonian}) for $I(t)\sim x_\theta^{\text out}\sim S_x$]. Note that this is just an illustration and the derivation needs to account for noise as well. Nevertheless, we can proceed in a similar way and expect the interaction as $\int_0^t h(z)S_x(t)\mathcal{F}[S_x(t-z)]dz$. For the linear feedback, $\mathcal{F}[S_x]=S_x$, this term resembles the long-range spin-spin interaction in space: here we have a long-range (i.e. long-memory) "interaction" of spins with themselves in the past. The "interaction length" is determined by $h(t)$.

Such a time-space analogy was successfully used in spin-boson model \cite{Leggett1987,LeHur2008, VojtaPRL2005,VojtaPh2006}, describing spins in a bosonic bath of nontrivial spectral function: $\omega^s$ for small frequencies [$s=1$ for Ohmic, $s<1$ ($s>1$) for sub-(super-)Ohmic bath, cf. \cite{Suppl}]. It was shown that a similar ``time-interaction'' term can be generated \cite{VojtaPRL2005,VojtaPh2006}. Moreover, an analogy with the spin chain and long-range interaction term in space $\sum_{i,j}S_iS_j/|r_i-r_j|^{s+1}$ was put forward and the break of the quantum-classical mapping was discussed \cite{VojtaPRL2005,VojtaPRL2012}. For $s=1$ a QPT of the Kosterlitz-Thouless type was found \cite{LeHur2008}, while QPTs for the sub-Ohmic baths are still under active research \cite{Plenio2011,Plenio2018}.

In bath problems, such a long-memory interaction can be obtained only asymptotically \cite{VojtaPRL2005,VojtaPh2006}. Moreover, arbitrarily tuning the spectral properties of quantum baths in a given system is challenging (cf. \cite{Leppakangas2018} for quantum simulations of the spin-boson model and \cite{Nokkala2016,Nokkala2018} for complex network approach). In contrast, the feedback response $h(t)$ can be implemented and varied naturally, as signals are processed digitally, opening paths for simulating various problems in a single setup. The function
\begin{eqnarray}\label{h}
h(t)=h(0)\left(\frac{t_0}{t+t_0}\right)^{s+1}
\end{eqnarray}
will correspond to the spatial Ising-type interaction. The instantaneous feedback with $h(t)\sim\delta(t)$ will lead to ``short-range in time'' $S_x^2$ term, as in the Lipkin-Meshkov-Glick (LMG) model \cite{Parkins2008} originating from nuclear physics. A sequence of amplitude-shaped time delays $h(t)\sim \sum_n \delta(t-nT)/n^{s+1}$ will enable studies of discrete time crystals \cite{Sacha2017,UedaTC2018,Demler2019,JakschNCom2019,JakschPRA2019} and Floquet engineering \cite{Eckardt2017} with long-range interaction $\sum_n S_x(t)S_x(t-nT)/n^{s+1}$, where the crystal period may be $T=2\pi/\omega_R$. This is in contrast to standard time crystals, where the parameter modulation is externally prescribed [e.g. periodic $g(t)$]. Here, the parameters are modulated depending on the system state (via $S_x$), i. e., self-consistently, as it happens in real materials e.g. with phonons. The “interaction in time” does not necessarily require the presence of standard atom-atom interaction in space. The global interaction is given by constant $h(t)$. The Dicke model can be restored even in the adiabatic limit by exponentially decaying and oscillating $h(t)$ mimicking a cavity. All such $h(t)$ can be realized separately or simultaneously to observe the competition between different interaction types. Our results do not rely on effective Hamiltonians \cite{Caballero2015a}. This discussion motivates us to use in further simulations $h(t)$, Eq. (\ref{h}), unusual in feedback control.

{\it Feedback-induced phase transition.-} We show the existence of FPT with controllable critical exponent by linearizing (\ref{Hamiltonian}) and assuming the linear feedback: $\mathcal{F}[x_\theta^{\text out}]=x_\theta^{\text out}$. Using the bosonization by Holstein-Primakoff representation \cite{DomokosPRA2016}:  $S_z=b^\dag b - N/2$, $S_-=\sqrt{N-b^\dag b}b$, $S_+=b^\dag\sqrt{N-b^\dag b}$, $S_x=(S_++S_-)/2$, we get
\begin{eqnarray}\label{Hamiltonian-lin}
H=\delta a^\dag a +\omega_R b^\dag b +(b^\dag+b) [g(a+a^\dag)+GI(t)].
\end{eqnarray}
The bosonic operator $b$ reflects linearized spin ($S_x\approx\sqrt{N}X$), and the matter quadrature is $X=(b^\dag+b)/2$. 

Weak measurements constitute a source of competition with unitary dynamics \cite{Mazzucchi2016PRA,Mazzucchi2016NJP, Mazzucchi2016SciRep,Kozlowski2016PRAnH}, which is well seen in quantum trajectories formalism \cite{Daley,Ruostekoski2014, Pedersen2014, Molmer2016PRA, VasilyevPRL2018, VasilyevPRA2018, Sherson2018PRA}, underlining the distinction between measurements and dissipation. Thus they can affect phase transitions, including the many-body ones \cite{Mazzucchi2016PRA,UedaCrit2016, Bason2018}. Feedback was mainly considered for stabilizing interesting states \cite{Sherson2015,Mazzucchi2016Opt,Sherson2016, Hush2013,Bouchoule2017,Hammerer2016PRA}. Here, we focus on the QPT it induces. In this formalism, the operator feedback signal $I(t)$ in Eq. (\ref{Hamiltonian-lin}) takes stochastic values $I_c(t)$ conditioned on a specific set (trajectory) of measurement results $\langle x_\theta\rangle_c(t)$ \cite{Wiseman}: $I_c(t)=\sqrt{2\kappa}\int_0^t h(t-z)[\sqrt{2\kappa}\langle x_\theta\rangle_c(z)+\xi(z)]dz$, where $\xi(t)$ is white noise, $\langle \xi(t+\tau)\xi(t)\rangle=\delta(\tau)$. The evolution of conditional density matrix $\rho_c$ is then given by \cite{Wiseman}: $d\rho_c=-i[H,\rho_c]dt+\mathcal{D}[a]\rho_c dt+\mathcal{H}[a]\rho_c dW$, where $\mathcal{D}[a]\rho_c=2\kappa[a\rho_c a^\dag-(a^\dag a\rho_c + \rho_c a^\dag a)/2]$, $\mathcal{H}[a]\rho_c=\sqrt{2\kappa}[a e^{-i\theta}\rho_c+\rho_c a^\dag e^{i\theta}-\text{Tr}(a e^{-i\theta}\rho_c+\rho_c a^\dag e^{i\theta})\rho_c]$, $dW=\xi dt$. In general, averaging such stochastic master equation over trajectories does not necessarily lead to the master equation for unconditional density matrix $\rho$ used to describe DPTs. 

Figure 2 compares trajectories for the spin quadrature $\langle X\rangle_c$ at various feedback constants $G$ and $s$ (\ref{h}). Crossing FPT critical point $G_\text{crit}$, the oscillatory solution changes to exponential growth. For large $s$ (nearly instant feedback), there is a frequency decrease before FPT and fast growth above it. For small $s$ (long memory), before FPT trajectories become noisier; the growth above it is slow. Note, that even though the trajectories are stochastic, their frequencies and growth rates are the same for all experimental realizations.

To get insight, we proceed with a minimal model necessary for FPT and adiabatically eliminate the light mode from Eq.~(\ref{Heis}): $a=(-2igX+f_a)/(\kappa+i\delta)$. This corresponds well to experiments \cite{EsslingerNat2010,EsslingerNature2016,Suppl}, where $\kappa$ ($\sim$ MHz) exceeds other variables ($\sim$ kHz). The Heisenberg equations for two matter quadratures then combine to a single equation describing matter dynamics:
\begin{eqnarray}\label{eqX}
\ddot{X}+\left(\omega^2_R  -\frac{4\omega_Rg^2\delta}{\kappa^2+\delta^2}\right)X - \nonumber\\    \frac{4\omega_RGg\kappa}{\kappa^2+\delta^2}C_\theta \int_0^t{ h(t-z)X(z)dz}=F(t),
\end{eqnarray}
where $C_\theta=\delta \cos \theta+\kappa \sin \theta$. Here the frequency shift is due to spin-light interaction, the last term originates from the feedback. The steady state of Eq.~(\ref{eqX}) is $\langle X\rangle=0$, which looses stability, if the feedback strength $G>G_\text{crit}$. 

Note, that oscillations below $G_\text{crit}$ are only visible at quantum trajectories for conditional $\langle X\rangle_c$ (Fig. 2). They are completely masked in the unconditional trivial solution $\langle X\rangle=0$. Thus, feedback can create macroscopic spin coherence $\langle X\rangle_c \ne 0$ at each single trajectory (experimental run) even below threshold. This is in contrast to dissipative systems, where the macroscopic coherence is attributed to $\langle X\rangle \ne 0$ above DPT threshold only.

\begin{figure}
\includegraphics[clip, trim=0.5cm 0.6cm 0.5cm 0.3cm, width=0.48\textwidth]{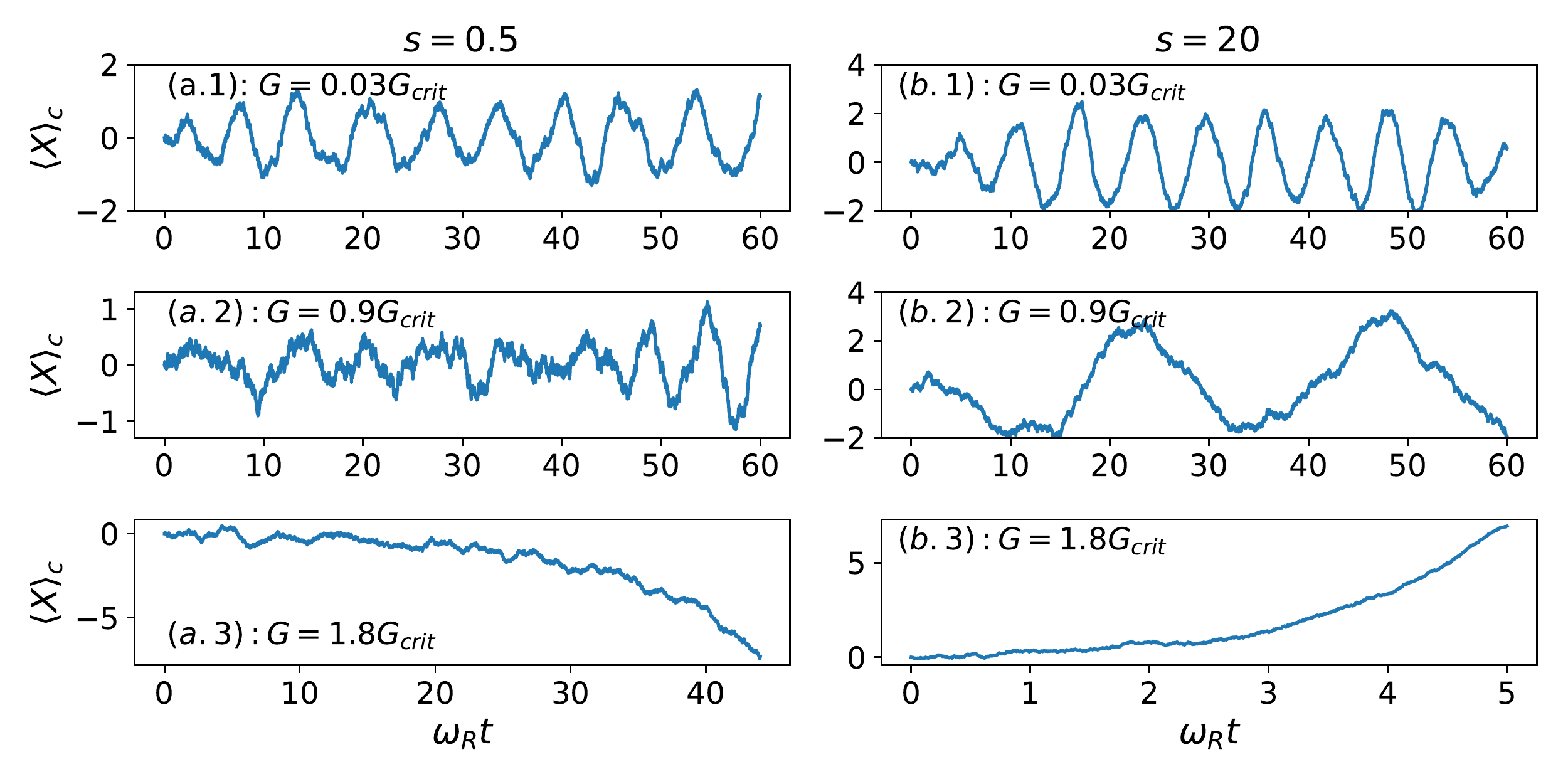}
\caption{\label{Fig2} Feedback-induced phase transition at a single trajectory. Conditional quadrature $\langle X\rangle_c$. For long-memory feedback [small $s=0.5$, panel (a)], approaching the transition at $G=G_\text{crit}$, the oscillatory trajectory becomes noisier and switches to slow growth. For fast feedback [large $s=20$, panel (b)], the oscillation frequency decreases (visualizing mode softening), and switches to fast growth. Even though the trajectories are stochastic, their frequencies and growth rates are the same for all experimental realizations. $g=\omega_R$, $\kappa=100\omega_R$, $\delta=\omega_R$, $\omega_Rt_0=1$. $h(0)=s$ gives the same $G_\text{crit}$ for all $h(t)$. Note the different scales on time axes. }
\end{figure}

The noise operator is $F(t)=-\omega_R[gf_a+G(\kappa-i\delta)e^{-i\theta}\int_0^t{h(t-z)f_a(z)dz/2}]/(\kappa+i\delta) +\text{H. c.}$ It has the following correlation function:
\begin{eqnarray} \label{tau}
\langle F(t+\tau) F(t)\rangle=
\frac{\omega_R^2\kappa}{2(\kappa^2+\delta^2)}\{4g^2\delta(\tau)+  \nonumber\\   
 G^2(\kappa^2+\delta^2)\int_0^t h(z)h(z+\tau)dz +  \nonumber\\
2gG\left[(\kappa-i\delta)e^{-i\theta}h(\tau)+ 
(\kappa+i\delta)e^{i\theta}h(-\tau)\right]\}.
\end{eqnarray}
We thus readily see how the feedback leads to the non-Markovian noise in spin dynamics.

Performing the Fourier transform of Eq.~(\ref{eqX}), one gets $D(\omega)\tilde{X}(\omega)=\tilde{F}(\omega)$, with the characteristic polynomial
\begin{eqnarray}\label{CharacterEq}
D(\omega)=\omega^2-\omega_R^2+\frac{4\omega_Rg^2\delta}{\kappa^2+\delta^2}+ 
\frac{4\omega_RGg\kappa}{\kappa^2+\delta^2}C_\theta H(\omega),
\end{eqnarray} 
where $\tilde{X}$, $\tilde{F}$, and $H(\omega)$ are transforms of $X$, $F$, and $h(t)$. The spectral noise correlation function is $\langle\tilde{F}(\omega)\tilde{F}(\omega')\rangle=S(\omega)\delta(\omega+\omega')$ with  
\begin{eqnarray}\label{SpNoise}
S(\omega)=\frac{\pi \omega_R^2\kappa}{\kappa^2+\delta^2}\left|2g+G(\kappa-i\delta)e^{-i\theta}H(\omega)\right|^2,
\end{eqnarray} 
whose frequency dependence again reflects the non-Markovian noise due to the feedback.

Even a simple feedback acting on spins leads to rich classical dynamics \cite{Kopylov2015}. Here we focus on the quantum case, but only for a simple type of phase transitions, where the eigenfrequency $\omega$ approaches zero \cite{Scarlatella2016} ("mode softening," visualized in quantum trajectories in Fig. 2). From the equation $D(\omega)=0$ we find the FPT critical point for the feedback strength: 
\begin{eqnarray}\label{Gcrit}
G_\text{crit}H(0)=\frac{1}{4g\kappa C_\theta}[\omega_R(\kappa^2+\delta^2)-4g^2\delta],
\end{eqnarray}
where $H(0)=\int_0^\infty{h(t)dt}$. Without feedback ($G=0$) this gives very large $g_\text{crit}$ for LMG and Dicke transitions \cite{DomokosPRL2015, DomokosPRA2016}. Thus, feedback can enable and control these transitions, even if they are unobtainable because of large decoherence $\kappa$ or small light-matter coupling $g$.

{\it Quantum fluctuations and critical exponent.-} We now turn to the quantum properties of FPT driven by the measurement-induced noise $F(t)$ (\ref{eqX}). While the mean-field solution is $\langle X\rangle=0$ below the critical point, $\langle X^2\rangle \ne 0$ exclusively due to the measurement fluctuations and can serve as an order parameter. From $D(\omega)\tilde{X}(\omega)=\tilde{F}(\omega)$ and noise correlations we get $\langle X(t+\tau)X(t)\rangle=\int_{-\infty}^\infty S(\omega)e^{i\omega\tau}/|D(\omega)|^2d\omega /(4\pi^2)$, giving  $\langle X^2\rangle$ for $\tau=0$.

To find the FPT critical exponent $\alpha$ we approximate the behavior near the transition point as $\langle X^2\rangle=A/|1-G/G_\text{crit}|^\alpha+B$, where $A,B=\text{const}$. Figure 3 demonstrates that the feedback can control the quantum phase transitions. Indeed, it does not only define the mean-field critical point (\ref{Gcrit}), but enables tuning the critical exponent as well. Varying the parameter $s$ of feedback response $h(t)$ (\ref{h}) allows one changing the critical exponent in a broad range. This corresponds to varying the length of effective spin-spin interaction mentioned above. For $h(t)$  (\ref{h}), its spectrum is expressed via the exponential integral $H(\omega)=h(0)t_0e^{-i\omega t_0}E_{s+1}(i\omega t_0)$. At small frequencies its imaginary part behaves as $\omega^s$ for $s<1$, resembling the spectral function of sub-Ohmic baths. For large $s$, $\alpha$ approaches unity, as $h(t)$ becomes fast and feedback becomes nearly instant  such as interactions in open LMG and Dicke models, where $\alpha=1$ \cite{DomokosPRL2015, DomokosPRA2016,Oztop2012}. 

\begin{figure}
\includegraphics[clip, trim=0cm 0.3cm 0.35cm 0.37cm, width=0.48\textwidth]{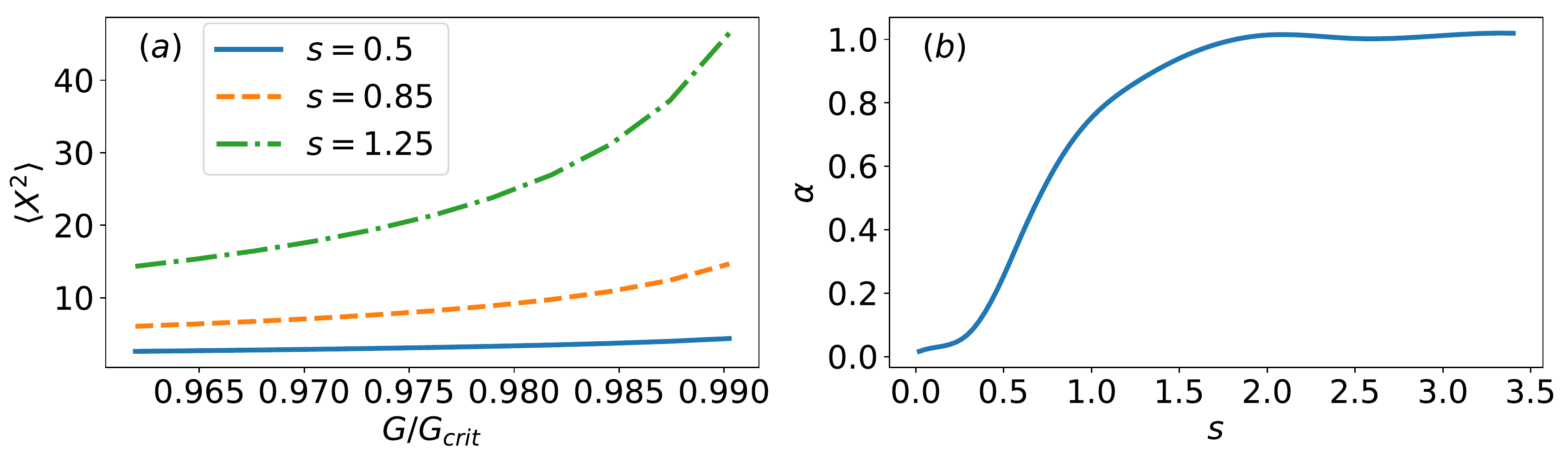}
\caption{\label{Fig3} Feedback control of critical exponent. (a) Growing fluctuations of unconditional matter quadrature $\langle X^2\rangle$ for various feedback exponents $s$. (b) Dependence of critical exponent $\alpha$ on feedback exponent $s$, proving opportunity for QPT control. $g=\omega_R$, $\kappa=100\omega_R$, $\delta=\omega_R$, $h(0)=s$, $\omega_Rt_0=1$.}
\end{figure}

Note that a decaying cavity is well known to produce the autonomous exponential feedback \cite{St-K2017} $h(t)=\exp(-\kappa' t)$ [$H(\omega)=1/(i\omega+\kappa')$] crucial in many fields (e.g. lasers, cavity cooling, optomechanics, etc.) Such a simple $H(\omega)$ is nevertheless insufficient to tune the critical exponent and measurement-based feedback is necessary. 

The linearized model describes FPT near the critical point, but it does not give new steady state. The spin nonlinearity can balance the system (cf.~\cite{Suppl}). However, the feedback with nonlinear $\mathcal{F}[x_\theta^{\text out}]$ can assure a new steady state even in a simple system of linear quantum dipoles (e.g. for far off-resonant scattering with negligible upper state population). It is thus the nonlinearity of the full hybrid quantum-classical system that is crucial.

{\it Relation to other models.-} Feedback control of QPTs enables simulating models similar to those for particle-bath interactions, e.g., spin-boson (SBM), Kondo, Caldeira-Leggett (CLM), quantum Brownian motion models (cf. \cite{Suppl}). They were applied to various systems from quantum magnets to cold atoms with various spectral functions \cite{breuer2002, Leggett1987, LeHur2008,  DomokosPRL2015,  DomokosPRA2016, Scarlatella2016, Plenio2011}. Creating a quantum simulator, which is able to model various baths in a single device, is challenging, and proposals include, e.g., coupling numerous cavities or creating complex networks simulating multimode baths \cite{Leppakangas2018, Nokkala2016, Nokkala2018}. In contrast, the feedback approach is more flexible as tuning $h(t)$ of a single classical loop is feasible. E.g., for BEC~\cite{Suppl}, the typical frequencies are in the kHz range, which is well below those of modern digital processors reaching GHz. Moreover, it can be readily extended for simulating broader class of quantum materials and qubits with nonlinear bath coupling \cite{ZhengPRB2018} and multiple baths \cite{VojtaPRL2012}.

The multi- (or large-) spin-boson models \cite{Anders2008, Winter2014, DomokosPRL2015,  DomokosPRA2016, Scarlatella2016} are based on Eq.~(\ref{Hamiltonian}) with sum over continuum of bosonic modes $a_i$ of frequencies $\delta_i$ distributed according to the spectral function $J(\omega)$ \cite{Suppl}. The feedback model reproduces exactly the form of bath dynamical equations for $S_{x,y,z}$ [cf. Eq.~(\ref{eqX}) for linearized, and \cite{Suppl} for nonlinear versions] if $\Im H(\omega)\sim J(\omega)-J(-\omega)$. The noise correlation function of linear CLM is $\langle\tilde{F}(\omega') \tilde{F}(\omega)\rangle=4\pi\omega_R^2J(\omega)\delta(\omega+\omega')$, whereas the feedback model contains $H(\omega)$ and additional light-noise term in Eq.~(\ref{SpNoise}). 

In bath models there is a delicate point of the frequency $\omega_R$ renormalization ("Lamb shift") \cite{Leggett1987, DomokosPRL2015,  DomokosPRA2016, Scarlatella2016}. It may lead to divergences and necessity to repair the model \cite{Ford1988}. The feedback approach is flexible. The frequency shift in Eq.~(\ref{CharacterEq}) is determined by $GH(0)=G\int_0^\infty h(t)dt$ and can be tuned and even made zero, if $h(t)$ changes sign.

In summary, we have shown that feedback does not only lead to phase transitions driven by quantum measurement fluctuations, but controls its critical exponent as well. It induces effects similar to those of quantum bath problems, allowing their realization in a single setup, and enables studies of time crystals and Floquet engineering with long-range (long-memory) interactions. The applications can also include control schemes for optical information processing \cite{JP2018}. Experiments can be based on quantum many-body gases in a cavity \cite{EsslingerNat2010, EsslingerNature2016, Hemmerich2015, LevPRL2018, Zimmermann2018, Mazzucchi2016Opt}, and circuit QED, where ultra-strong coupling has been obtained \cite{Yoshihara2017, Forn2017} or effective spins can be considered \cite{Leppakangas2018, HuardUltraS2018,Ustinov2017}. Feedback methods can be extended by, e.g., measuring several outputs \cite{Hammerer2016PRA, Hacohen2016,HuardNCom2018} (enabling simulations of qubits and multi-bath SBMs \cite{VojtaPRL2012} with nonlinear couplings \cite{ZhengPRB2018}) or various many-body atomic \cite{Elliott2015, Mazzucchi2016SciRep, Kozlowski2017, Mazzucchi2016Opt} or molecular \cite{MekhovLP2013} variables. 

{\it Note.-} After the acceptance of our letter, the first experiment, where our predictions can be tested was reported in Ref. \cite{EsslingerFeedback}.

\begin{acknowledgments} 
We thank Ph. Joyez, A. Murani, and D. Esteve for stimulating discussions. Figures are prepared using MS PowerPoint. Support by RSF (17-19-01097),  RFBR (18-02-01095), DGAPA-UNAM (IN109619), CONACYT-Mexico (A1-S-30934), EPSRC (EP/I004394/1), UPSay (d'Alembert Chair).  
\end{acknowledgments} 

\bibliographystyle{apsrev4-2}

%

\end{document}


\title
{Feedback-Induced Quantum Phase Transitions Using Weak Measurements - Supplemental Material}

\author{D. A. Ivanov}
\affiliation{Department of Physics, St. Petersburg State University, St. Petersburg, Russia}
\author{T. Yu. Ivanova}
\affiliation{Department of Physics, St. Petersburg State University, St. Petersburg, Russia}
\author{S. F. Caballero-Benitez}
\affiliation{Instituto de F\'isica, Universidad Nacional Aut\'onoma de M\'exico, Ciudad de M\'exico, M\'exico}
\author{I. B. Mekhov}
\affiliation{Department of Physics, St. Petersburg State University, St. Petersburg, Russia}
\affiliation{Department of Physics, University of Oxford, Oxford, United Kingdom}
\affiliation{SPEC, CEA, CNRS, Universit\'{e} Paris-Saclay, CEA Saclay, Gif-sur-Yvette, France}

\maketitle

\renewcommand{\thefigure}{S\arabic{figure}} 
\renewcommand\theequation{S\arabic{equation}}

\section{Feedback control of quantum phase transitions in ultracold gases}

The model presented in the paper can be realized using light scattering from the Bose-Einstein condensate (BEC). Two spin levels will correspond to two motional states of ultracold atoms. A very high degree of the light-matter interaction control has been achieved in several systems, where BEC was trapped in an optical cavity, and the Dicke and other supersolid-like phase transitions were obtained \cite{EsslingerNat2010,LevPRL2018, Zimmermann2018}. Experiments now include strongly correlated bosons in an optical lattice inside a cavity \cite{EsslingerNature2016, Hemmerich2015} and related works without a cavity \cite{Kettrle-selforg2017}. Here we propose one realization and underline that setups can be flexible and extendable for other configurations as well.

\begin{figure}
\includegraphics[clip, trim=0cm 14cm 13cm 0cm, width=0.5\textwidth]{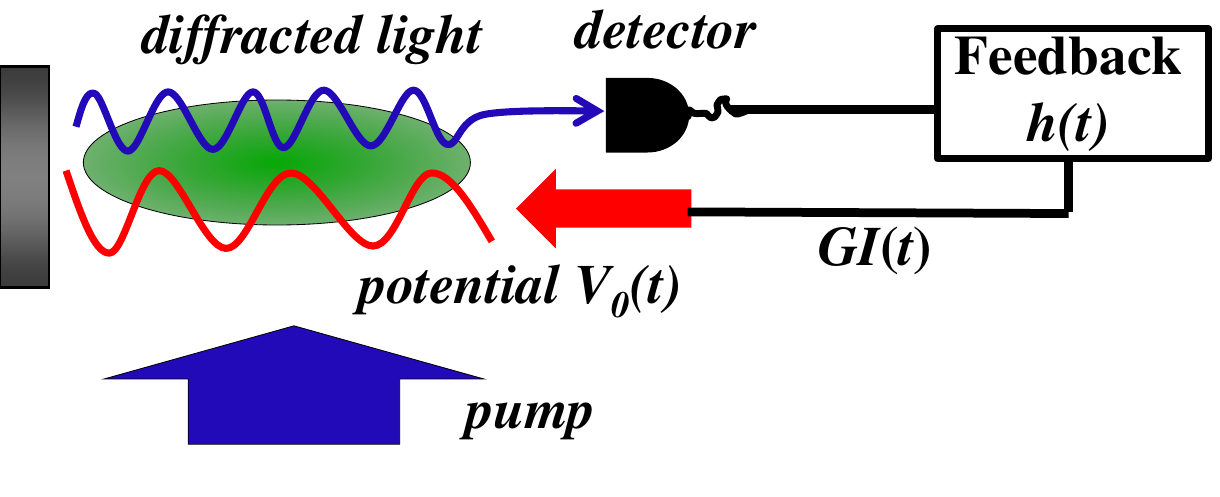}
\caption{\label{SupplSetup} Setup. A BEC is illuminated by the transverse pump, the scattered (diffracted) light is detected, feedback acts on the system via the change of the external periodic potential depth $V_0(t)$. Feedback provides the non-Markovianity, nonlinearity, and noise, necessary for the controllable quantum phase transition. Importantly, the feedback response $h(t)$ is tunable.} 
\end{figure}

We consider a BEC, elongated in the $x$ direction, illuminated by the pump (Fig. S1). The quadrature of scattered light is measured and used for the feedback signal. The feedback is provided by the external trapping potential in the form of a standing wave, whose depth is varied according to the feedback signal: $V_0(t)\cos^2 k_0x$ ($k_0$ is the wave vector of laser beam creating the potential). The many-body Hamiltonian has a form (cf. review \cite{Mekhov2012}, we work in the units, where $\hbar=1$):
\begin{eqnarray}\label{Hamiltonian}
H=\sum_l \omega_l a^\dag_l a_l +\int_0^L \Psi^\dag(x)H_{a1}\Psi(x)dx+\sum_l \zeta_l(a^\dag_l+a_l),
\end{eqnarray}
where $a_l$ are the annihilation operators of light modes of frequencies $\omega_l$ interacting with a BEC, $\zeta_l$ are the pumps of these modes (if they are shaped by cavities), $H_{a1}$ is the single-atom Hamiltonian, $\Psi(x)$ is the atom-field operator, and $L$ is the BEC length. For the far off-resonant interaction, $H_{a1}$ is determined by the interference terms between the light fields present \cite{Mekhov2012}:
\begin{eqnarray}\label{Hamiltonian-a1}
H_{a1}=\frac{p^2}{2m_a}+V_0(t)\cos^2k_0x+\frac{1}{\Delta_a}\sum_{l,m} g_l u^*_l(x)a^\dag_l  g_m u_m(x)a_m,
\end{eqnarray}
where the first term is atomic kinetic energy operator ($p^2=d^2/dx^2$, $m_a$ is the atom mass), $\Delta_a$ is the detuning between light modes and atomic transition frequency, $g_{l,m}$ is the light-matter coupling constants. $u_{l,m}$ are the geometrical mode functions of light waves, which can describe pumping and scattering at any angles to the BEC axis \cite{MekhovPRL2007,MekhovPRA2007}, which maybe convenient depending on the specific experimental realization. 

Here, to strongly simplify the consideration, we select the following geometry of light modes (cf. Fig. S1). The pump is orthogonal to the BEC axis, thus its mode function is constant along $x$ (can be chosen as $u(x)=1$); its amplitude $a_\text{pump}$ is considered as a c-number. A single scattered light mode $a_1$ is non-negligible along $x$ direction, and its mode function is $u_1(x)=\cos k_1x$, where $k_1$ is the mode wave vector. There is no direct mode pumping, $\zeta_1=0$. The condition $k_0=k_1/2$ assures the maximal scattering of light into the mode $a_1$ (diffraction maximum). Thus, the pump diffracts into the mode $a_1$ from the atomic distribution. Thus, the wavelength of feedback field should be twice the mode wavelength. The matter-field operator can be decomposed in two modes \cite{DomokosPRL2015, DomokosPRA2016}:
\begin{eqnarray}\label{Psi}
\Psi(x)=\frac{1}{\sqrt{L}}c_0+\sqrt{\frac{2}{L}}c_1\cos k_1x,
\end{eqnarray}
where $c_{0,1}$ are the annihilation operators of the atomic waves with momenta $0$ and $k_1$ ($c^\dag_0c_0+c^\dag_1c_1=N$, $N$ is the atom number). Substituting Eqs. (\ref{Psi}) and (\ref{Hamiltonian-a1}) in Eq. (\ref{Hamiltonian}) and neglecting in a standard way several terms \cite{DomokosPRL2010,EsslingerNat2010,DomokosPRL2015, DomokosPRA2016} (which however may appear to be important under specific conditions and thus enrich physics even further), we get the Hamiltonian (1) in the main text
\begin{eqnarray}\label{Hamiltonian-FB}
H=\delta a_1^\dag a_1 +\omega_R S_z +\frac{2}{\sqrt{N}}S_x [g(a_1+a_1^\dag)+GI(t)]
\end{eqnarray}
 with the following parameters: $\delta=\omega_1-\omega_\text{pump}+Ng_1^2/(2\Delta_a)$ (the detuning between mode and pump including the dispersive shift), $\omega_R=k_1^2/2m_a$ is the recoil frequency, $g=\Omega_\text{pump} g_1\sqrt{N/2}/\Delta_a$ with $\Omega_\text{pump}=g_\text{pump}a_\text{pump}$ being the pump Rabi frequency. The feedback signal is $GI(t)=\sqrt{N/8}V_0(t)$. The spin operators are $S_x=(c^\dag_1c_0+c_1c^\dag_0)/2$, $S_y=(c^\dag_1c_0-c_1c^\dag_0)/(2i)$, $S_z=(c^\dag_1c_1-c^\dag_0c_0)/2$. The main characteristic frequency of this system is the recoil frequency that for BEC experiments is $\omega_R=2\pi \cdot 4$kHz. This makes the feedback control feasible, as the modern digital procossing of the feedback signal can be much faster (up to the GHz values). Other experimental parameters such as $\delta$ and $g$ (depending on the pump amplitude) can be tuned in the broad range and, in particular, be close to kHz values. The cavity decay rate in Ref.  \cite{EsslingerNat2010} is $\kappa=2\pi\cdot 1.3$ MHz, which is much greater than kHz making the adiabatic elimination of the light mode to work very well.
 
There are other configurations relevant to our work. For example, for the BEC in a cavity setup \cite{EsslingerNat2010}, instead of creating the external potential $V_0$ with $k_0=k_1/2$, one can inject the feedback signal directly through the cavity mirror as the pump $\zeta_1(t)$. In this case, an additional laser with doubled wavelength is not necessary. While the Hamiltonian Eq. (\ref{Hamiltonian-FB}) will be somewhat different, after the adiabatic elimination of light mode, the equation for the matter quadrature operator $X$ [Eq. (5) in the main text] will be the same with $\zeta_1(t)\sim I(t)$.

Another possibility is to use atoms tightly trapped in optical lattices \cite{EsslingerNature2016, Hemmerich2015}. In this case, instead of expanding $\Psi(x)$ in the momentum space, a more appropriate approach is to expand it in the coordinate space using localized Wannier functions \cite{Mekhov2012}. The spin operators will then be represented by sums of on-site atom operators $S_x\sim \sum_i A_i n_i$, or the bond operators representing the matter-wave interference between neighboring sites $S_x\sim \sum_i A_i b^\dag_i b_j$ \cite{Caballero2015,CaballeroNJP2016}. For example, the effective spin can correspond to the atom number difference between odd and even sites [for $A_i=(-1)^i$] \cite{Mazzucchi2016PRA}, represent the magnetization \cite{Mazzucchi2016PRA} or staggered magnetization \cite{Mazzucchi2016SciRep} of fermions, etc. Such a strong flexibility in choosing the geometrical combination of many-body variables combined in the effective spin operator enables defining macroscopic modes of matter fields \cite{Elliott2015} and assures the competition between the long-range (but structured in space with a short period comparable to that of the lattice) light-induced interactions and short-range atom-atom interactions and tunneling on a lattice \cite{Mazzucchi2016PRA}. This will open the opportunities for the competition between the nontrivial feedback-induced interactions and many-body atomic interactions.

We belive that our proposal will extend the studies in the field of time crystals \cite{Sacha2017,UedaTC2018,Demler2019,JakschNCom2019,JakschPRA2019} and Floquet engineering \cite{Eckardt2017}. The time crystals is a recently proposed notion, where phenomena studied previously in space (e.g. spin chains, etc.) are now studied in time. Typically, the system is subject to the external periodic modulation of a parameter [e.g. periodic $g(t)$], which is considered as creating a “lattice in time”. Our approach makes possible introducing the effective “interaction in time.” This makes the modulation in the system not prescribed, but depending on the state of the system (via $S_x$). This resembles a true lattice in space with the interaction between particles. In other words, our model enables not only creating a “lattice in time,” but introducing the tunable “interaction in time” to such a lattice (without the necessity of having the standard particle-particle interaction in space).
 
 {\it Note.-} After the acceptance of our letter, the first experiment, where our predictions can be tested was reported in Ref. \cite{EsslingerFeedback}.

\section{Bath models}

The Hamiltonian of the spin-boson model at zero temperature \cite{Leppakangas2018} extended for $N$ spins is 
\begin{eqnarray}\label{SB-Hamiltonian}
H=\sum_i\delta_i a_i^\dag a_i +\omega_R S_z +\frac{2}{\sqrt{N}}S_x \sum_i g_i(a_i^\dag+a_i).
\end{eqnarray}
It describes the interaction of spins with many (continuum) bosonic modes $a_i$ of frequencies $\delta_i$. The corresponding Heisenberg-Langevin equations are then given by
\begin{eqnarray} \label{HL-Res}
\dot{a_i}=-i\delta_i a_i -i\frac{2g_i}{\sqrt{N}}S_x, \nonumber\\
\dot{S_x}=-\omega_R S_y, \nonumber\\
\dot{S_y}=\omega_RS_x - \frac{2}{\sqrt{N}} S_z\sum_i g_i(a_i^\dag+a_i), \nonumber\\
\dot{S_z}=\frac{2}{\sqrt{N}}S_y\sum_ig_i(a_i^\dag+a_i).
\end{eqnarray}

Two equations for $S_x$ and $S_y$ can be joined: $\ddot{S_x} =-\omega_R^2S_x+2\omega_RS_z/\sqrt{N}\sum_i g_i(a_i^\dag+a_i)$. The modes $a_i$ can be formally found from the first Eq. (\ref{HL-Res}): $a_i(t) =a_i(0)e^{-i\delta_it}-2ig_i/\sqrt{N}\int_0^tS_x(\tau)e^{-i\delta_i(t-\tau)}d\tau$. This can be combined to $\sum_i g_i(a_i^\dag+a_i)=-4\int_0^\tau\beta(t-\tau)S_x(\tau)d\tau/\sqrt{N}-F_b(t)/\omega_R$, where
\begin{eqnarray}
\beta(t)=\sum_i g_i^2(\delta_i)\sin(\delta_i t)=\frac{1}{\pi}\int_0^{\infty}J(\omega)\sin\omega t d\omega, 
\end{eqnarray}
and the bath spectral function is
\begin{eqnarray}\label{J}
J(\omega)=\pi\sum_i g^2_i(\delta_i)\delta(\omega-\delta_i).
\end{eqnarray}
The noise operator $F_b(t)$ is determined by random initial values of the bosonic mode operators $a_i(0)$: $F_b(t)=-\omega_R\sum_i g_i [a_i(0)e^{-i\delta_it}+a_i^\dag(0)e^{i\delta_it}]$.

At the level of Heisenberg-Langevin equations, the linearized system can be obtained by assuming $S_z=-N/2$, $S_x=\sqrt{N}X$, and $S_y=\sqrt{N}Y$. Alternatively, it can be obtained from the Hamiltonian (\ref{SB-Hamiltonian}) using the Holstein-Primakoff representation as explained in the main text. The bosonized Hamiltonian then reads
\begin{eqnarray}
H=\sum_i\delta_i a_i^\dag a_i +\omega_R b^\dag b +(b^\dag+b) \sum_i g_i(a_i^\dag+a_i),
\end{eqnarray}
while the linearized equation for $S_x$ is reduced to the equation for the particle quadrature $X$:
\begin{eqnarray}\label{X-Res}
\ddot{X}+\omega_R^2 X- 4\omega_R\int_0^t \beta(t-z)X(z)dz = F_b(t).
\end{eqnarray}

Such Hamiltonian and operator equation correspond to the quadrature-quadrature coupling model and differs from the Caldeira-Leggett model only by the renormalized spin frequency $\omega_R$ \cite{breuer2002}. They describe the quantum Brownian motion as well \cite{breuer2002}. Equation (\ref {X-Res}) has a structure identical to Eq. (5) obtained for the feedback model in the main text. The additional frequency shift in the case of feedback can be compensated either by modifying the spin frequency, or by adding the $\delta(t)$-term in the feedback response function $h(t)$. 

The bath spectral function $J(\omega)$ (\ref{J}) is usually approximated as $J(\omega)=\kappa_R(\omega/\omega_c)^sP_c(\omega)$, where $\omega_c$ is the cut-off frequency and $P_c(\omega)$ is the cut-off function. For $s=1$ the bath is called Ohmic, for $s<1$ it is sub-Ohmic, and for $s>1$ it is super-Ohmic. This corresponds to the bath response function asymptotically behaving as $1/t^{s+1}$ for large times \cite{LeHur2008}.  

Taking the Fourier transform of the differential equation (\ref{X-Res}) one gets
\begin{eqnarray}
\left(\omega^2-\omega_R^2+4\omega_RB(\omega)\right)X(\omega)=-\tilde{F_b}(\omega),
\end{eqnarray}
where $B(\omega)$ is the Fourier transform of $\beta(t)$. The spectral noise correlation function reads
\begin{eqnarray}
\langle\tilde{F_b}(\omega') \tilde{F_b}(\omega)\rangle=4\pi\omega_R^2J(\omega)\delta(\omega+\omega').
\end{eqnarray}

The time noise correlation function is
\begin{eqnarray}
\langle F_b(t+\tau) F_b(t)\rangle=\frac{\omega_R^2}{\pi}\int_0^\infty J(\omega)e^{-i\omega\tau} d\omega =
\frac{\omega_R^2}{\pi}\left[\int_0^\infty J(\omega)\cos\omega\tau d\omega - i\int_0^\infty J(\omega)\sin\omega\tau d\omega \right].
\end{eqnarray}

A more standard way to write the differential equation is not via $\beta(t)$, but via $\gamma(t)$:
\begin{eqnarray}
\ddot{X}+\omega_R^2\left(1-\frac{2\gamma(0)}{\omega_R}\right) X+ 2\omega_R\int_0^t \gamma(t-z)\dot{X}(z)dz = F_b(t),
\end{eqnarray}
where $\dot{\gamma}(t)=-2\beta(t)$,
\begin{eqnarray}
\gamma(t)=\frac{2}{\pi}\int_0^{\infty}\frac{J(\omega)}{\omega}\cos\omega\tau d\omega. 
\end{eqnarray}
The frequency shift can be incorporated in the renormalized spin frequency. For the Ohmic bath, $\gamma(t)\sim \delta(t)$, and  the differential equation is reduced to that for a damped harmonic oscillator.

\section{Feedback-induced phase transition with nonlinear spins}

In the main text, we presented the properties of the phase transition using the linearized model with Hamiltonian (4). Such a linear model cannot give us the value of the stationary state above the critical point. Here we show the results of numerical simulations for a single nonlinear spin. The original Hamiltonian (1) is 
\begin{eqnarray}\label{Hamiltonian-FB-2}
H=\delta a^\dag a +\omega_R S_z +\frac{2}{\sqrt{N}}S_x [g(a+a^\dag)+GI(t)].
\end{eqnarray}
The Heisenberg-Langevin equations are then given by
\begin{eqnarray} \label{HL-FB}
\dot{a}=-i\delta a -i\frac{2g}{\sqrt{N}}S_x-\kappa a +f_a, \nonumber\\
\dot{S_x}=-\omega_R S_y, \nonumber\\
\dot{S_y}=\omega_RS_x - \frac{2}{\sqrt{N}} [g(a+a^\dag)+GI(t)]S_z, \nonumber\\
\dot{S_z}=\frac{2}{\sqrt{N}}[g(a+a^\dag)+GI(t)]S_y.
\end{eqnarray}

Figure 2 shows the results of numerical simulations for the expectation value of the component $\langle S_x\rangle$ of a single spin ($N=1$). It shows the phase transition with $\langle S_x\rangle=0$ below $G_\text{crit}$ and  $\langle S_x\rangle \ne 0$ above the critical point. This is similar to the spin-boson model, where $\langle S_x\rangle \ne 0$ corresponds to the localization, while $\langle S_x\rangle = 0$ corresponds to the delocalized phase.

\begin{figure}
\includegraphics[clip, trim=0cm 0cm 0cm 0cm, width=0.5\textwidth]{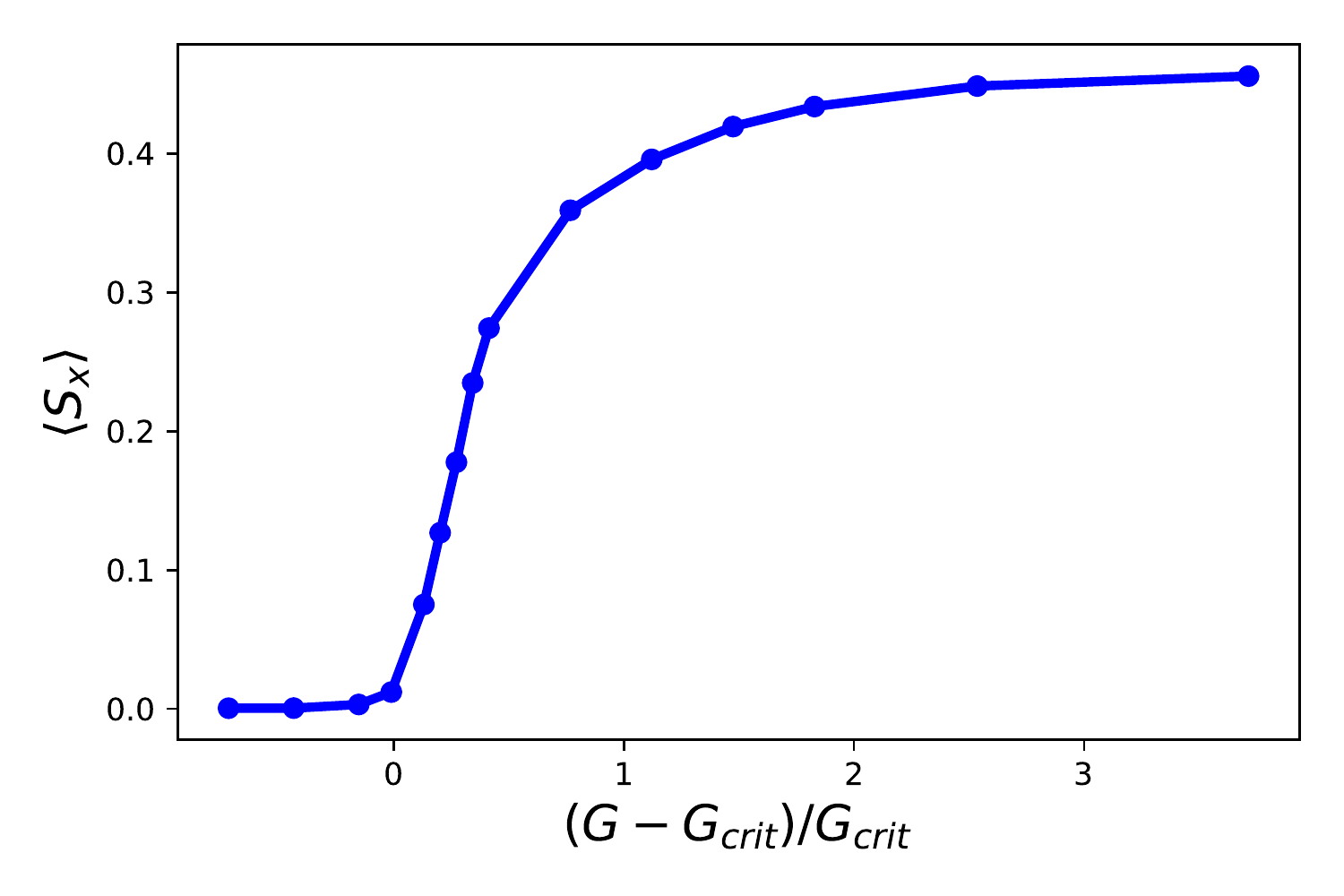}
\caption{\label{Suppl-Spin} Feedback-induced phase transition for a single spin. $\delta=\omega_R$, $g=0.1 \omega_R$,  $h(0)=s$,  $s=1$, $\kappa=10\omega_R$, $\omega_Rt_0=1$.} 
\end{figure}

Such solutions can be obtained by calculating the conditional expectation values $\langle S_x(t)\rangle_c$ and then averaging them over multiple quantum trajectories, which gives the stationary unconditional expectation value $\langle S_x\rangle$. In general, the classical measurement and feedback loop produce a single quantum trajectory. Averaging over many trajectories reproduces the results of quantum Heisenberg-Langevin equations (the equivalence between the quantum trajectory and quantum Heisenberg-Langevin approaches is demonstrated for feedback e.g. in Ref. \cite{Wiseman}). In turn, similar Heisenberg-Langevin approach describes the interaction of a particle with quantum baths as well.

Similarly to the bath problem, the equations for $S_x$ and $S_y$ can be joined: $\ddot{S_x} =-\omega_R^2S_x+2\omega_RS_z/\sqrt{N}[g(a^\dag+a)+GI(t)]$. The light mode can be adiabatically eliminated from the first equation (\ref{HL-FB}) to give $g(a^\dag+a)+GI(t)=-[4g^2\delta S_x + 4g\kappa G C_\theta \int_0^t h(t-\tau)S_x(\tau)d\tau]/[\sqrt{N}(\kappa^2+\delta^2)]-F(t)/\omega_R$, where the notations and noise correlations are given in the main text. These expressions have the same structure as those in the bath models. Various methods to treat the quantum Heisenberg-Langevin equations have been developed in quantum optics \cite{Davidovich,Andreev,ScullyBook}. The linearized equation for $S_x$ then reduces to Eq. (5) of the main text for the particle quadrature $X$ ($S_x=\sqrt{N}X$):
\begin{eqnarray}
\ddot{X}+\left(\omega^2_R  -\frac{4\omega_Rg^2\delta}{\kappa^2+\delta^2}\right)X -     \frac{4\omega_RGg\kappa}{\kappa^2+\delta^2}C_\theta \int_0^t{ h(t-z)X(z)dz}=F(t),
\end{eqnarray}
which again has the same structure as the equation for $X$ in the bath model. The additional frequency shift in the case of feedback can be compensated either by modifying the spin frequency, or by adding the $\delta(t)$-term in the feedback response function $h(t)$. 

Finally, we would like to comment on the role of limited detector efficiency $\eta<1$. In the Heisenberg-Langevin approach, it can be taken into account by introducing an additional light noise corresponding to undetected photons leaked from the cavity, while the noise $f_a$ will still correspond to the detected photons (cf. Ref. \cite{Habibi}). The characteristic equation (7) in the main text will then take exactly the same form with $G$ replaced by $\sqrt{\eta}G$. Therefore, concerning the position of the critical point, the limited efficiency can be compensated by the increase of the feedback coefficient $G$. Indeed, when the detector efficiency approaches zero  ($\eta=0$) such that it can not be compensated in a real system, the role of feedback (detected photons) vanishes, and the feedback-induced phase transition reduces to a standard dissipative phase transition due to the undetected photons (i.e. the dissipation).

\bibliographystyle{apsrev4-2}

%